\documentclass[10pt,twocolumn,twoside] {IEEEtran}
\usepackage{booktabs}  
\usepackage{multirow}
\usepackage{bm}
\usepackage{array}
\usepackage{graphicx}
\usepackage{epstopdf}
\usepackage{subfigure}
\usepackage{amssymb}
\usepackage{amsmath}
\usepackage{makecell}
\usepackage{color}
\usepackage{algorithmic}
\usepackage{algorithm}

\def\x{{\bf x}}

\def\0{{\bf 0}}
\def\1{{\bf 1}}

\usepackage{todonotes}
\usepackage[pagebackref=false,breaklinks=true,letterpaper=true,colorlinks,bookmarks=false]{hyperref}
\usepackage{cite}

\usepackage[pagebackref=false,breaklinks=true,letterpaper=true,bookmarks=false]{hyperref}
\usepackage[square, comma, sort&compress, numbers]{natbib}

\graphicspath{{figs/}}

\emergencystretch=\hsize
\begin{document}

\title{Augmentation-based Unsupervised Cross-Domain Functional MRI Adaptation for Major Depressive Disorder Identification}

\author{Yunling Ma, Chaojun Zhang, Xiaochuan Wang, Qianqian Wang, Liang Cao, Limei Zhang, Mingxia Liu

}

\markboth{}
{Ma~\MakeLowercase{\textit{et al.}}: Augmentation-based Unsupervised Cross-Domain fMRI Adaptation}
	
\maketitle

\begin{abstract}
Major depressive disorder (MDD) is a common mental disorder that typically affects a person's mood, cognition, behavior, and physical health. 
Resting-state functional magnetic resonance imaging (rs-fMRI) data are widely used for computer-aided diagnosis of MDD.
While multi-site fMRI data can provide more data for training reliable diagnostic models, significant cross-site data heterogeneity would result in poor model generalizability.
Many domain adaptation methods are designed to reduce the distributional differences between sites to some extent, but usually ignore overfitting problem of the model on the source domain.
Intuitively, target data augmentation can alleviate the overfitting problem by forcing the model to learn more generalized features and reduce the dependence on source domain data.
In this work, we propose a new augmentation-based unsupervised cross-domain fMRI adaptation (AUFA) framework for automatic diagnosis of MDD. The AUFA consists of 1) a graph representation learning module for extracting rs-fMRI features with spatial attention, 
2) a domain adaptation module for feature alignment between source and target data,
3) an augmentation-based self-optimization module for  alleviating model overfitting on the source domain,  
and 4) a classification module. 
Experimental results on 1,089 subjects suggest that AUFA outperforms several state-of-the-art methods in MDD identification. 
Our approach not only reduces data heterogeneity between different sites, 
but also localizes disease-related functional connectivity abnormalities and provides interpretability for the model.
\end{abstract}

\begin{IEEEkeywords}
Resting-state functional MRI; Domain adaptation; Multi-site data; Major depressive disorder
\end{IEEEkeywords}
\IEEEpeerreviewmaketitle

\section{Introduction}
\begin{figure*}[!tp]
 	\setlength{\abovecaptionskip}{-14pt} 
 	\setlength{\belowcaptionskip}{-2pt} 
 	\setlength\abovedisplayskip{-2pt}
 	\setlength\belowdisplayskip{-2pt}
 	\centering
 	\includegraphics[width=1\textwidth]{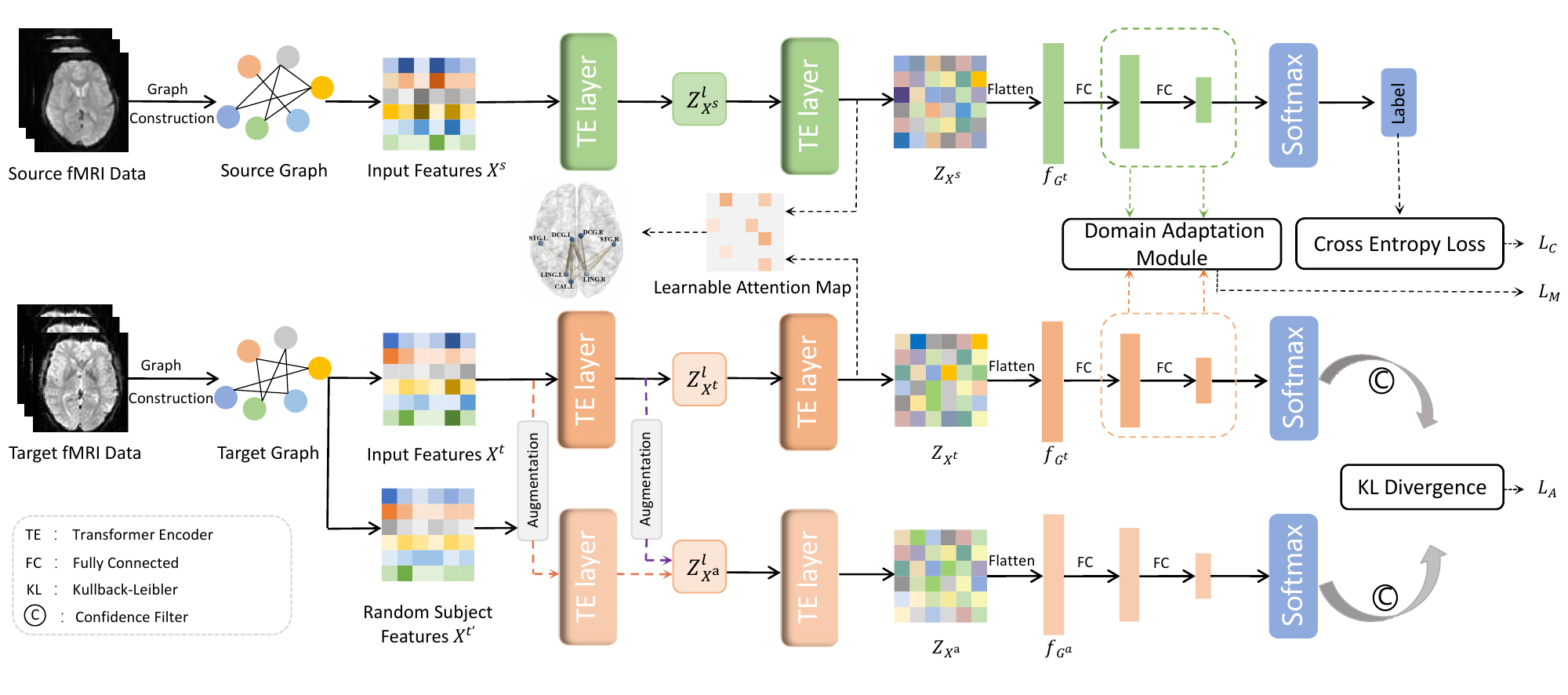}
 	\caption{Overview of our augmentation-based unsupervised cross-domain fMRI adaptation (AUFA) framework, including (1) fMRI feature extraction for brain functional connectivity networks using a Transformer module, (2) domain adaptation module for feature alignment between source and target domains, (3) augmentation-based self-optimization module, and (4) classification module. 
   The data from the target domain are fed
   into the model that has been jointly optimized by $L_{C}$, $L_{M}$ and $L_{A}$ for direct prediction.
  }
 	\label{framework}
 \end{figure*}

Major depressive disorder (MDD) is a common mental health disorder, leading to symptoms such as profound sadness, self-denial, feelings of helplessness, and impaired cognitive function, thus significantly impacting the patient's daily life and work~\cite{proudman2021growing}. 
Currently, the diagnosis of MDD mainly relies on the patient's symptom assessment, physical examination, and psychological scales, which still involves subjectivity and risk of misdiagnosis~\cite{pan2018diagnosis,zheng2016metabolite}. 
Therefore, precise identification of MDD is crucial for clinicians to develop prompt intervention plans.

Resting-state functional magnetic resonance imaging (rs-fMRI) is used as a noninvasive, nonradioactive technique to assess changes in neural activity over time in the whole brain.
It is extensively utilized in computer-aided analysis of brain disorders using learning-based methods~\cite{gray2020multimodal,sui2018multimodal}.
Currently, machine learning and deep learning techniques primarily focus on utilizing functional connectivity networks (FCNs) to capture the intricate interplay among regions of interest (ROIs) within the brain~\cite{sun2021estimating}. These networks are constructed from blood-oxygen-level dependent (BOLD) time series data obtained from rs-fMRI~\cite{ma2023multi}.
Traditional machine learning methods often extract handcrafted features based on FCN for brain disease detection. However, these methods require prior knowledge of FCN~\cite{liu2021plant}, and shallow features usually result in poor performance in classification tasks.

With advances in deep learning, various methods employ graph neural networks to automatically extract high-level feature representations from fMRI time series for automatic diagnosis of MDD~\cite{zhang2023multi,wang2024leveraging}. 
For example, Li \emph{et al}.~\cite{pitsik2023topology} learned topological features from brain functional connectivity constructed based on fMRI data, and utilized a graph neural network (GNN) to identify MDD. 
Jun \emph{et al}.~\cite{jun2020identifying} integrated valid connectivity features from rs-fMRI with non-imaging phenotypic features, 
followed by training a graph convolutional network (GCN) on the integrated features.
However, these methods only use a single site from the MDD dataset for experiments, which does not guarantee the generalization ability of the model. 
Multi-site rs-fMRI data provide more data for learning-based methods to train reliable models~\cite{wang2022multi,fang2023addressing}. 
However, fMRI acquired from different sites often exhibit significant heterogeneity due to equipment differences and variations in data acquisition conditions~\cite{yamashita2019harmonization}, which presents a challenge for diagnosing brain diseases.

Domain adaptation helps reduce the distributional gap between domains~\cite{farahani2021brief}. However, 
when there is a large discrepancy between the source and target domains, along with the size limitations of fMRI datasets, overfitting on the source domain is more likely to occur~\cite{sun2022safe}.
That is, the model is too fine-grained in capturing the source domain features and ignores the generalization of the universal laws. 
Intuitively, augmenting the target domain data can improve the model's ability to generalize to the target domain and avoid over-reliance on specific source domain data~\cite{zhou2022domain,alzubaidi2021novel}.

To this end, we develop a new augmentation-based unsupervised cross-domain fMRI adaptation (AUFA) framework for automated diagnosis of MDD.
As shown in Fig.~\ref{framework}, we first construct FCNs for source/target domain subjects based on fMRI data. 
Then, we employ the Transformer encoder to learn fMRI features with spatial attention.  
At the same time, the maximum mean
difference constraint is introduced to align the fMRI features learned between domains, aiming for the Transformer encoder to learn shared features between domains. 
Furthermore, to enhance the model's generalization ability to the target domain, we optimize it by augmenting target domain data to prevent overfitting on the source domain.
The AUFA significantly reduces the data distribution differences across sites through the supervision of source domain labels and adaptation of data between domains. 
Experiments on rs-fMRI data from 1,089 subjects indicate the excellent performance of AUFA in 
identifying MDD compared to other competing methods.

The major contributions of this work are summarized as follows.
\begin{itemize}
\item A novel augmentation-based unsupervised cross-domain fMRI adaptation (AUFA) framework is proposed for automated diagnosis with rs-fMRI data.
\item The AUFA introduces the maximum mean difference constraint to align fMRI features from different sites, which can significantly reduce the discrepancy of data distribution between source and target domains.
\item The self-optimization module based on augmentation in AUFA helps prevent the overfitting of the model in the source domain and improves the generalization ability in the target domain.
\item AUFA can effectively capture the attention relationship between different ROIs to automatically learn fMRI features and obtain discriminative functional connectivity.
\item Cross-domain disease detection experiments at 5 fMRI imaging sites demonstrate the superiority of our method over several advanced methods.
\end{itemize}

The remainder of this paper is structured as follows.
A concise overview of relevant studies is provided in Section~\ref{S2}.
Section~\ref{S3} introduces the data acquisition and pre-processing. 
Section~\ref{S4} introduces the details of the proposed AUFA. 
We describe the experimental settings, comparison methods, and experimental results in Section~\ref{S5}. 
Section~\ref{S6} discusses the primary components of the proposed AUFA and lists several limitations
in the present work.
Finally, the paper is summarized in Section~\ref{S7}.

\section{Related Work}
\label{S2}
\subsection{Learning-based Methods for Functional MRI Analysis}
Deep learning, as a powerful machine learning method, has been widely used in fMRI analysis~\cite{yin2022deep,liu2018applications}. 
Deep learning models exhibit superior capabilities in handling high-dimensional, nonlinear, and complex fMRI data, showcasing better modeling and generalization performance compared to traditional machine learning methods~\cite{wen2018deep}.
Graph neural networks, as a class of deep learning algorithms dedicated to processing graph data, have also been widely used for the analysis of FCN based on fMRI, such as graph convolutional networks (GCNs)~\cite{kipf2016semi}. 
GCNs learn node representations by aggregating features of nodes and their neighbors through convolutional operations. 
Yao \emph{et al}.~\cite{yao2021mutual} developed a mutual multi-scale triplet GCN for diagnosing brain diseases, analyzing both brain functional and structural connectivity.
Noman \emph{et al}.~\cite{noman2024graph} designed a new GCN-based graphical autoencoder (GAE) framework for MDD detection.
However, these methods usually ignore the extent to which different brain regions contribute to downstream tasks.
Transformer is an attention-based deep learning model originally proposed for natural language processing tasks (\emph{e.g}., machine translation)~\cite{vaswani2017attention}. 
It introduces a self-attention mechanism that enables modeling the dependencies between different positions in an input sequence. In recent years, the Transformer model has also been applied as a powerful sequence modeling tool for capturing attention relationships between brain regions in fMRI data analysis~\cite{asadi2023transformer,bedel2023bolt}. 
In this work, we attempt to leverage the Transformer in unsupervised domain adaptation for multi-site fMRI feature extraction. Considering that the strong backbone of Transformer increases the risk of overfitting in the source domain, we augment the target domain data and improve the model generalization through a self-optimization module.

\subsection{Domain Adaptation of Functional MRI Data}
The multi-site fMRI data greatly increased the sample size for the model training. 
However, due to distributional differences among data from different sites, the training data may not precisely represent the distribution of the testing data, leading to a degradation in model performance~\cite{lee2021meta,yamashita2019harmonization}. The domain adaptation approach is an important strategy for solving this problem.
Currently, domain adaptation methods can be categorized into three groups based on the availability of target domain labels~\cite{guan2021domain}:
(1) supervised domain adaptation~\cite{motiian2017unified}, (2) Semi-supervised domain adaptation~\cite{li2021learning}, and (3)unsupervised domain adaptation~\cite{patel2015visual}.
In this work, we focus on unsupervised domain adaptation, where only unlabeled target data can be used to train the adaptive model. 

Recently, domain adaptation methods have been extensively used in the field of computer vision~\cite{zhao2020review}.
For example, Zhou \emph{et al}.~\cite{zhou2021unsupervised} proposed a new Adversarial Distribution Adaptation Network (ADAN) to 
simultaneously minimize global and local distributional differences between domains, facilitating the learning of domain-invariant representations.
Li \emph{et al}.~\cite{li2022unsupervised} proposed a Progressive Adaptation of Subspace (PAS) method to gradually obtain reliable pseudo-labels for model training. 
Liu \emph{et al}.~\cite{liu2018structure} developed a new unsupervised domain adaptation method that preserves the data distribution of the source domain to guide the learning of target data structure through a semi-supervised clustering approach.
However, multi-site adaptation studies for fMRI time series are still limited. 
Some studies based on multi-site fMRI data have simply combined fMRI data from different sites for training~\cite{gallo2021thalamic}, \emph{i.e}., assuming that they have the same data distribution, ignoring data heterogeneity between sites. 
Even though some studies attempt to reduce the distributional differences of fMRI data through domain adaptation methods, they usually ignore the problem that the model is likely to be overfitted on the source domain. 
For example, Wang \emph{et al}.~\cite{wang2020unsupervised} proposed an unsupervised graph domain adaptation network for autism spectrum detection, comprising three components: graph isomorphism encoder, progressive feature alignment, and unsupervised information matrix regularization. 
Lee \emph{et al}.~\cite{lee2021meta} designed a novel domain generalization meta-learning model to identify brain diseases based on rs-fMRI.
The aim of this paper is to study unsupervised domain adaptation for multi-site fMRI data from two perspectives, \emph{i.e}., reducing inter-domain distributional differences and mitigating the risk of model overfitting.

\begin{table}[!tbp]
\setlength{\belowcaptionskip}{-4pt}
\setlength{\abovecaptionskip}{-1pt}
\setlength\abovedisplayskip{-1pt}
\setlength\belowdisplayskip{-1pt}
\renewcommand\arraystretch{1}
\centering
\caption{{\color{black}Phenotypic information of 1089 subjects from REST-meta-MDD consortium~\cite{yan2019reduced}. MDD: Major depressive disorder; NC: Normal control; M/F: Male/Female.}}
 \scriptsize
 \centering
 \begin{center}
  \begin{tabular*}{0.46\textwidth}{@{\extracolsep{\fill}} l|cccc}
   \toprule
   ~~Site ID & Category & Subject~ &  Gender (M/F)&   Age (Years)   \\
   \hline
   ~~~~\multirow{2}{*} {20}
   &MDD & 282  &$99/183$& $38.74\pm13.63$\\  
   &NC &251  &$87/164$& $39.64\pm15.84$\\
   \hline
   ~~~~\multirow{2}{*} {21}
   &MDD & 86  &$38/48$& $34.71\pm12.56$\\  
   &NC &70  &$31/39$& $36.13\pm12.55$\\
   \hline
   ~~~~\multirow{2}{*} {25}
   &MDD & 89  &$21/68$& $65.60\pm6.72$\\  
   &NC &63  &$29/34$& $69.63\pm5.82$\\
   \hline
   ~~~~~\multirow{2}{*} {1}
   &MDD & 74  &$31/43$& $31.72\pm8.14$\\  
   &NC &74  &$32/42$& $31.80\pm8.93$\\
   \hline
   ~~~~~\multirow{2}{*} {9}
   &MDD & 50  &$23/27$& $28.10\pm8.74$\\  
   &NC &50  &$31/19$& $28.92\pm8.50$\\
   \bottomrule
  \end{tabular*}
 \end{center}
 \label{MDD}
\end{table}

\section{Materials and Data Preprocessing}
\label{S3}

\subsection{Data Acquisition}
This work involved rs-fMRI data from the top five largest sites in the imaging center of the REST-meta-MDD consortium~\cite{yan2019reduced}, including SITE 20, SITE 21, SITE 25, SITE 1, and SITE 9.
There were a total of 1,089 participants across the five sites, including 581 MDD patients and 508 normal controls (NCs). 
The rs-fMRI data from SITE 20, SITE 21 and SITE 1 were acquired by Siemens Verio 3T with the following parameters: echo time (TE)$/$repetition time (TR) $ = 30\,ms/2, 000\,ms$, flip angle $= 90$.
Particularly, at SITE 20, 
the thickness$/$gap was $3.0\,mm/1.0\,mm$ with $32$ slice; at SITE 21, the thickness$/$gap was $3.5\,mm/0.7\,mm$ with $32$ slice, and at SITE 1, the thickness$/$gap was $4.0\,mm/0.8\,mm$ with $32$ slice.
The rs-fMRI data from SITE 25 were obtained by Siemens Verio 3T with the following parameters: TE$/$TR $ = 25\,ms/2, 000\,ms$, flip angle $ = 90$, thickness$/$gap $= 4.0\,mm/0\,mm$, slice number $ = 36$.
The rs-fMRI data for SITE 9 were obtained by GE Discovery MR750 3.0T with the following parameters: TE$/$TR $ = 25\,ms/2, 000\,ms$, flip angle $ = 90$, thickness$/$gap = $3.0\,mm/1.0\,mm$, slice number $ = 35$.
Table~\ref{MDD} reports more detailed demographic information.

\begin{figure*}[!tp]
 	\setlength{\abovecaptionskip}{-1pt}
 	\setlength{\belowcaptionskip}{-2pt} 
 	\setlength\abovedisplayskip{-2pt}
 	\setlength\belowdisplayskip{-2pt}
 	\centering
 	\includegraphics[width=1\textwidth]{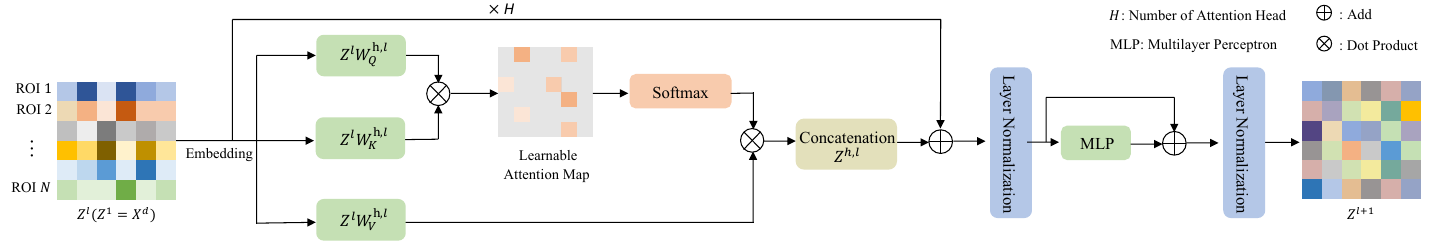}
 	\caption{Graph representation learning module in AUFA that learns new features with spatial attention through multiple Transformer encoder layers.
  }
 	\label{representation learning}
 \end{figure*}

\subsection{Image Pre-Processing}
The initial rs-fMRI data were pre-processed using the Data Processing Assistant for Resting-State fMRI (DPARSF)~\cite{yan2010dparsf} software, including the following steps: (1) discard the first ten time points, (2) head motion correction, (3) space normalization and smoothing, (4) interference signal regression, (5) band-pass filtering for noise reduction. Finally, we extracted an average rs-fMRI time series for 116 predefined ROI based on Automated Anatomical Labeling (AAL)~\cite{tzourio2002automated} atlas.

\section{Proposed Method}
\label{S4}

\subsection{Problem Formulation}
In this work, we focus on utilizing cross-domain rs-fMRI data for the MDD identification task, aiming to solve an interesting, realistic, but challenging unsupervised domain adaptation problem, \emph{i.e}., 
Utilizing knowledge of the source domain to help predict completely unlabeled target domain data.
We first introduce some notation that will be used in this study.
Let $X^{d}$ be the input (graphs constructed based on fMRI data) space and $Y^{d}$ be the target (label) space, where $d$ identifies either the source domain $s$ or the target domain $t$. 
The source domain is defined as $S$ = $\{X^{s}, Y^{s}\}$, containing the feature set $X^{s}$ and the corresponding label set $Y^{s}$. The target domain is defined accordingly as $T$ = $\{X^{t}, Y^{t}\}$ with feature set $X^{t}$ and label set $Y^{t}$. 
Even though the two domains have the same task (they share the label space $Y^{s}$ = $Y^{t}$), their data distribution is somewhat different, \emph{i.e}., $P^{s}(X)\neq P^{t}(X)$. 
In the framework of unsupervised domain adaptation, the challenge is how to learn a feature extractor $\Psi:X\rightarrow C$ that leverages the abundant labeling information from the source domain. 
This extractor maps the initial input to a feature space 
$C$, aiming to reduce the data distribution between the source and target domains, \emph{i.e}., 
$P^{s}(\Psi(X))\approx P^{t}(\Psi(X))$.

\subsection{Method}
We propose an augmentation-based unsupervised cross-domain fMRI adaptation (AUFA) framework for MDD identification.
As shown in Fig.~\ref{framework}, the AUFA comprises four main components: 
(1) graph representation learning for extracting rs-fMRI features based on Transformer from the constructed brain FCNs, (2) domain adaptation for feature alignment, (3) augmentation-based self-optimization module, and (4) classification.

\subsubsection{Graph Representation Learning Module}
\label{1}
This module includes two parts: graph construction and Transformer-based graph representation learning, with details introduced below. 

\textbf{Graph Construction.}
The brain is a complex neural network system, wherein different ROIs are responsible for respective functions and interact with each other to coordinate cognitive processes and behaviors.
The FCN constructed by analyzing the fMRI time series relationship between different ROIs can effectively capture the complex relationship in the brain.
Each subject's FCN can be modeled as an undirected graph $G$ = $(V, E)$, where $V$ denotes the node set, representing different ROIs, and $E$ denotes the edge set, representing the strength of functional connections between ROI pairs. 
FCNs are widely employed in computer-aided detection of MDD due to their ability to capture the topology of the brain.

In this work, we first construct FCNs for subjects from the source and target domains. 
We utilize Pearson correlation (PC) to calculate correlations of rs-fMRI time series, establishing functional connections between ROI pairs. The functional connection $w_{ij}\in\ [-1,1]$ between the $i$-th ROI and the $j$-th ROI is defined as:
\begin{equation}
\footnotesize
\begin{aligned}
w^{d}_{ij}=\frac{(x^{d}_{i}-\bar{x}^{d}_{i})^{T}(x^{d}_{j}-\bar{x}^{d}_{j})}{\sqrt{(x^{d}_{i}-\bar{x}^{d}_{i})^{T}(x^{d}_{i}-\bar{x}^{d}_{i})}\sqrt{(x^{d}_{j}-\bar{x}^{d}_{j})^{T}(x^{d}_{j}-\bar{x}^{d}_{j})}},\\
\end{aligned}
\label{Eq1}
\end{equation}
where $\bar{x}^{d}_{i}$ denotes the mean vector of the BOLD signal ${x}^{d}_{i}$ corresponding to the $i$-th ROI, with $d$ representing the source or target domain. Thus, for each subject, we obtain the functional connectivity matrix $X^{d}$, where $X^{d}_{ij}$=$w^{d}_{ij}$, and utilize the constructed $X^{d}\in R^{N\times N}$ as the initial graph feature for the source and target domains.

\textbf{Transformer-based Graph Representation Learning.}
To accurately capture the specific contributions of individual brain regions and learn more discriminative features, we adopted the Transformer encoder for graph representation learning. 
This strategy utilizes the Transformer's self-attention mechanism to weigh the importance of different ROIs and learn the corresponding attention coefficients. 
Subsequently, based on the attention weights, the feature information of global nodes is aggregated and updated. Thus, it facilitates a deeper understanding of the structure of complex brain networks. 
More details are introduced in the following.

We take the node features $X^{d}$  as input, and model the relationship between each node  
and all other nodes through the self-attention mechanism.
The attention scores between nodes are calculated and the corresponding node representations are weighted and aggregated.
As shown in Fig.~\ref{representation learning}, define $Z^{l} (Z^{1}=X^{d})$ to be the graph representation of some subject in layer $l$. Taking the self-attention module in the $h$-th header as an example, this component is computed as follows:
\begin{equation}
\footnotesize
\begin{aligned}
Z^{h,l}&=Attention(Z^{l}W^{h,l}_{Q},Z^{l}W^{h,l}_K,Z^{l}W^{h,l}_V)=A^{h,l}Z^{l}W^{h,l}_V,
\end{aligned}
\label{Eq2}
\end{equation}
where $A^{h,l}=softmax(\frac{Z^{l}W^{h,l}_{Q}(Z^{l}W^{h,l}_K)^{T}}{\sqrt{D_{k}}})\in R^{N\times N}$ is the matrix of attention scores among all nodes in the $l$-th layer of the $h$-th header, 
$W^{h,l}_{Q}$, $W^{h,l}_{K}$, $W^{h,l}_{V}$ is the matrix of learnable linear transformations exerted on the node features, and $D_{k}$ is the dimension of $Z^{l}W^{h,l}_K$.

To capture more comprehensive information about the graph data, 
we employ a multi-head self-attention mechanism for parallel computation. The input $X^{d}$ is passed to $H$ different self-attention modules and $H$ output matrices $Z$ are computed and aggregated.
The aggregation operation can be represented as:
\begin{equation}
\small
\begin{aligned}
\bar{Z^{l}}=N(W(C(Z^{1,l},Z^{2,l},...,Z^{H,l}))),
\end{aligned}
\label{Eq3}
\end{equation}
where $C$ is the concatenation operation, $W$ denotes linear transform, and $N$ denotes layer normalization.

After feature learning through self-attention layers
, a further nonlinear transformation is performed by a feed-forward neural network. The feed-forward neural network is a two-layer fully connected neural network with the following formulation:
\begin{equation}
\small
\begin{aligned}
{Z^{l+1}}=N(RELU(\bar{Z}^{l}W_{1}+b_{1})W_{2}+b_{2}),
\end{aligned}
\label{Eq4}
\end{equation}
where $W_{1}$, $W_{2}$, $b_{1}$, $b_{2}$ are the learnable weights and biases. 
Thus, a new feature representation $Z_{X^{d}}$ of $X^{d}$ is obtained through the learning of multiple Transformer layers.
To conduct the subsequent graph classification task, the $Z_{X^{d}}$ are flattened to obtain the final graph-level representation $f_{X^{d}}$.

\subsubsection{Domain Adaptation Module}
To reduce domain offset, 
a maximum mean difference constraint is utilized to align the learned features across different domains.
As shown in  Fig.~\ref{framework}, we learn the graph-level feature $f_{X^{d}}$ of the source and target domains through the Transformer module learning. 
The $f_{X^{d}}$ are then fed into two fully connected layers, which are finally predicted by a softmax layer. 

Here, considering that the low-level features contain limited semantic information, we only align the graph-level feature mappings between domains, represented as follows:
\begin{equation}
\small
\begin{aligned}
L_{M} = \frac{1}{B^2} \sum_{l=1}^{L} (f_{s}^{l} - f_{t}^{l})^T (f_{s}^{l} - f_{t}^{l}), 
\end{aligned}
\label{Eq5}
\end{equation}
where 
$f_{s}^{l}$ represents the feature of the $l$-th layer in the source domain, $f_{t}^{l}$ represents the feature of the $l$-th layer in the target domain, $B$ is the batch size, and $L$ denotes the number of layers in the fully connected layer.
Eq.~(\ref{Eq5}) measures the discrepancy in feature distribution between the two domains. By minimizing this equation, we penalize the features with different distributions across domains, 
compelling the Transformer feature extractor to learn shareable feature representations across the two domains.

\subsubsection{Augmentation-based Self-optimization Module}
In Section~\ref{1}, we utilize a Transformer encoder to learn the graph representation for each subject. 
However, the model is prone to overfitting due to the limited training data and the large number of parameters included in the multilayer self-attention mechanism and feed-forward neural network.
In the unsupervised domain adaptation setting, the risk of the model overfitting the source domain data is significantly heightened as a result of the model's reliance solely on supervision from the source domain labels. 
Therefore, we reduce the risk of model overfitting on the source domain data by augmenting the target domain data to improve model generalizability.

To enhance model robustness, we randomly select the input features of the $l$-th layer Transformer encoder for augmentation in the target domain.
Given the initial graph feature $X^{t}$ to be augmented, we utilize another randomly selected subject $X^{t'}$ from the target domain to add an offset to $X^{t}$ to achieve augmentation. 
Let $Z^{l}_{X^{t}}$, $Z^{l}_{X^{t'}}$ denote the input features of $X^{t}$ and $X^{t'}$ in layer $l$, respectively. 
The features of subject $X^{a}$ in the $l$-th layer after augmentation can be expressed as:
\begin{equation}
\small
\begin{aligned}
Z^{l}_{X^{a}}=Z^{l}_{X^{t}}+\gamma(Z^{l}_{X^{t'}}-{Z}^{l}_{X^{t}}),
\end{aligned}
\label{Eq6}
\end{equation}
where $\gamma$ is a scalar that is used to adjust the degree of offset to the target domain features $X^{t}$. 
In fact, such an augmentation operation is equivalent to imposing regularization constraints on the model. 
Note that by randomly selecting the $l$-th layer for augmentation, the learning of subsequent parameters from the $(l+1)$-th layer to the last layer will be affected, thus achieving a multilayer regularization constraint on the model.
As shown in Fig.~\ref{framework}, $Z^{l}_{X^{t}}$ is augmented to the new target domain data $Z^{l}_{X^{a}}$  after the perturbation of $Z^{l}_{X^{t'}}$ in the $l$-th layer. Then, the new feature representations $Z_{X^{t}}$ and $Z_{X^{a}}$ are obtained through the subsequent learning of the Transformer layer.
We calculate the  prediction probabilities of $Z_{X^{t}}$ and $Z_{X^{a}}$ by performing the same procedure as  mentioned (\emph{i.e.}, flatten, fully connected layers and softmax), and
denoted as $P_{X^{t}}$, $P_{X^{a}}$.
Then, we compute the Kullback-Leibler ($KL$) divergence between $P_{X^{t}}$ and $P_{X^{a}}$ to minimize it:
\begin{equation}
\small
\begin{aligned}
KL(P_{i}||P_{j}) = \sum P_{i}~log\frac{P_{i}}{P_{j}},
\end{aligned}
\label{Eq7}
\end{equation}
where $i,j \in \{X^{t},X^{a}\}, i \neq j$. 
We update the model parameters according to the joint effect of $P_{X^{t}}$ and $P_{X^{a}}$ to avoid excessive gradients generated by any single probability.

It is important to mention that we lack label information to determine the reliability of predictions about the target domain.
Therefore, a confidence filter is added to round off samples corresponding to prediction probabilities with low confidence. The set of remaining target domain data can be represented as:
\begin{equation}
\small
\begin{aligned}
F=\{x\in D|max(P_{i})\textgreater \epsilon\},
\end{aligned}
\label{Eq8}
\end{equation}
where $D = \{X^{t}\cup X^{a}\}$ and $\epsilon$ is a threshold for filtering the prediction probability.
Thus, the final augmentation-based self-optimization loss is defined as:
\begin{equation}
\small
\begin{aligned}
L_{A}=KL(F(P_{i})||P_{j}),
\end{aligned}
\label{Eq9}
\end{equation}
where $i,j \in \{X^{t},X^{a}\}$, $i \neq j$. 

\subsubsection{Classification}
The graph-level representation of each subject is fed into two fully connected layers for classification, and the prediction probability vector is obtained through a softmax layer. 
The number of neurons in the first and second layers was $4096$ and $2$, respectively.
The whole model is trained using a joint loss function as the optimization objective, which comprises three components: (1) cross-entropy loss $L_{C}$ measures the discrepancy between the model's prediction for source domain data and the real label to ensure the model's classification performance in source domain, (2) maximum average difference loss $L_{M}$ is used to align the learned inter-domain features, (3) augmentation-based self-optimization loss $L_{A}$ is utilized to further optimize the model and mitigate overfitting of the model on the source domain data. Therefore, the joint loss function can be represented as:
\begin{equation}
\small
\begin{aligned}
L=L_{C}+\lambda_{1}L_{M}+\lambda_{2}L_{A},
\end{aligned}
\label{Eq10}
\end{equation}
where $\lambda_{1}$, $\lambda_{2}$ are hyperparameters. 

\subsection{Implementation Details}
\label{S3_A}
The AUFA framework is implemented on Pytorch using NVIDIA Quadro RTX 6000.
The model is trained in two stages. During the first stage, AUFA is trained in a supervised manner based on the source domain to narrow down the search space for parameters. 
The pre-trained AUFA is further optimized in the second stage by minimizing the joint loss function Eq.~(\ref{Eq10}) based on labeled source samples and unlabeled target samples.
We set the number of layers in the Transformer encoder to $2$ and the number of attention heads to $4$. 
Hyperparameters $\lambda_{1}$ and $\lambda_{2}$ are both $1$.
We optimize the AUFA model using the Adam algorithm with a learning rate of $0.00001$, a training epoch of $45$, and a batch size of $32$.

\section{Experiments}
\label{S5}
\subsection{Experimental Settings}
In this work, we design 4 cross-site fMRI prediction tasks: SITE 20 $\rightarrow$ SITE 21, SITE 20 $\rightarrow$ SITE 25, SITE 20 $\rightarrow$ SITE 1, SITE 20 $\rightarrow$ SITE 9. 
The site in front of the arrow is the source domain and behind the arrow is the target domain.
The source domain data contains full category labels and the target domain data does not have any label information. 
We employ source and target domain samples for training and predict the category to which the target domain samples belong based on the trained model parameters.
The model was trained with 5 random initializations, and the mean and standard deviation results were recorded.
The model was evaluated using 5 metrics, including (1) accuracy, (2) precision, (3) recall, (4) area under the receiver operating characteristic curve (AUC), and (5) F1-score.

\begin{table*}[!tbp]
\setlength{\belowcaptionskip}{-4pt}
\setlength{\abovecaptionskip}{-1pt}
\setlength\abovedisplayskip{-1pt}
\setlength\belowdisplayskip{-2pt}
\centering
\footnotesize
\renewcommand{\arraystretch}{1.1}
\caption{
Experimental results (mean$\pm$standard deviation) of the eight methods in the MDD vs. NC classification task, with the best results highlighted in bold.}
	\scriptsize
	\centering
	\begin{center}
    \begin{tabular*}{1\textwidth}{@{\extracolsep{\fill}} l|c  cc cc |ccccc }
			\toprule
		    ~ Method     & Accuracy  & Precision           & Recall           & AUC           & F1-score   & Accuracy  & Precision           & Recall           & AUC           & F1-score          \\
			\hline
           \multicolumn{6}{c|}{SITE 20 $\rightarrow$ SITE 21}  & \multicolumn{5}{c}{SITE 20 $\rightarrow$ SITE 25} \\
         \hline
    ~BC-SVM  & 0.51 & 0.55 & 0.63 & 0.49 & 0.59 
            & 0.51  & 0.57 & 0.66 & 0.49 & 0.61 \\
    ~LE-SVM  & 0.52 & 0.56  & 0.64 & 0.52 &0.59 
            & 0.53 & 0.61 & 0.57 & 0.50 & 0.59 \\
    ~MNSFS   & 0.48  & 0.56 & 0.29 & 0.49 & 0.38 
            & 0.49 & 0.58 & 0.45 & 0.52 & 0.51 \\
    ~GIN    & 0.52$\pm$0.04 & 0.57$\pm$0.03  & 
         0.55$\pm$0.13 & 0.51$\pm$0.03  & 0.55$\pm$0.08  
        & 0.50$\pm$0.04 & 0.60$\pm$0.05 & 0.50$\pm$0.12 & 0.48$\pm$0.06 & 0.53$\pm$0.07\\
    ~ST-GCN  & 0.52$\pm$0.05 & 0.56$\pm$0.04 & 
        0.60$\pm$0.18 & 0.49$\pm$0.04 & 0.57$\pm$0.08   
        & 0.57$\pm$0.01  & 0.62$\pm$0.02 & \textbf{0.71$\pm$0.09} & 0.55$\pm$0.03 & 0.65$\pm$0.03   \\
    ~CORAL   & 0.54$\pm$0.02 & 0.59$\pm$0.01 & 
        0.58$\pm$0.02 & 0.56$\pm$0.03  & 0.59$\pm$0.01  & 0.57$\pm$0.02  & 0.67$\pm$0.04  &0.56$\pm$0.01 & \textbf{0.60$\pm$0.03}  & 0.61$\pm$0.02  \\
    ~DANN    & 0.56$\pm$0.03    & 0.62$\pm$0.06 & 0.56$\pm$0.13  & 0.58$\pm$0.06 & 0.57$\pm$0.04  
        & 0.54$\pm$0.09  & 0.60$\pm$0.07 & 0.58$\pm$0.28 & 0.55$\pm$0.08 & 0.57$\pm$0.15  \\
    ~AUFA~(Ours)  &\textbf{0.63$\pm$0.03} & 
      \textbf{0.66$\pm$0.03}  & \textbf{0.73$\pm$0.06} & \textbf{0.62$\pm$0.04} & \textbf{0.69$\pm$0.03}   
      & \textbf{0.60$\pm$0.04} & \textbf{0.68$\pm$0.04} & 0.64$\pm$0.03  & 0.59$\pm$0.04 & \textbf{0.66$\pm$0.03}   \\
        \hline
         \multicolumn{6}{c|}{SITE 20 $\rightarrow$ SITE 1}  & \multicolumn{5}{c}{SITE 20 $\rightarrow$ SITE 9} \\
         \hline
    ~BC-SVM  & 0.49 & 0.49 & 0.58  & 0.49 & 0.53 
            & 0.51  & 0.51 & 0.74  & 0.46 & 0.60 \\
    ~LE-SVM  & 0.51 & 0.52 & 0.43  & 0.53 & 0.47 
            & 0.48  & 0.48 & 0.54  & 0.46 & 0.51  \\
    ~MNSFS   & 0.50 & 0.50 & 0.23 & 0.45 & 0.31 
            & 0.48  & 0.47 & 0.34 & 0.44 & 0.40 \\
    ~GIN    & 0.54$\pm$0.05 & 0.55$\pm$0.05 & 
          0.63$\pm$0.22 & 0.59$\pm$0.03 & 0.57$\pm$0.07
          & 0.56$\pm$0.03  & 0.56$\pm$0.07 & 0.70$\pm$0.20 & 0.56$\pm$0.04 & 0.61$\pm$0.06\\
    ~ST-GCN & 0.49$\pm$0.02 & 0.50$\pm$0.01 & 
        \textbf{0.84$\pm$0.10} & 0.54$\pm$0.02 & 0.62$\pm$0.03   
        & 0.51$\pm$0.04 & 0.52$\pm$0.08  & 0.66$\pm$0.03 & 0.48$\pm$0.06 & 0.51$\pm$0.22   \\
    ~CORAL  & 0.59$\pm$0.06  & \textbf{0.61$\pm$0.05}  & 
         0.56$\pm$0.05 & 0.61$\pm$0.06  & 0.59$\pm$0.05  
         & 0.55$\pm$0.03 & 0.57$\pm$0.02 & 0.57$\pm$0.04 & 0.60$\pm$0.05  & 0.57$\pm$0.03  \\
    ~DANN   & 0.59$\pm$0.03  & 0.59$\pm$0.03 & 
         0.60$\pm$0.08  & 0.60$\pm$0.03 & 0.59$\pm$0.04  & 0.55$\pm$0.05 & 0.54$\pm$0.04 & 0.70$\pm$0.23 & 0.59$\pm$0.04 & 0.60$\pm$0.09  \\
    ~AUFA~(Ours)  & \textbf{0.61$\pm$0.03} & 
         0.60$\pm$0.03  & 0.73$\pm$0.03
         & \textbf{0.63$\pm$0.02} & \textbf{0.66$\pm$0.03}    
         & \textbf{0.62$\pm$0.05} & \textbf{0.61$\pm$0.04}  & \textbf{0.78$\pm$0.05}  & \textbf{0.68$\pm$0.04} & \textbf{0.68$\pm$0.03}   \\
			\bottomrule
		\end{tabular*}
	\end{center}
	\label{results}
\end{table*}

\begin{figure*}[!tp]
	\setlength{\abovecaptionskip}{-2pt}
	\setlength{\belowcaptionskip}{-2pt}
	\setlength\abovedisplayskip{-1pt}
	\setlength\belowdisplayskip{-1pt}
	\centering
    \includegraphics[width=0.98\textwidth]{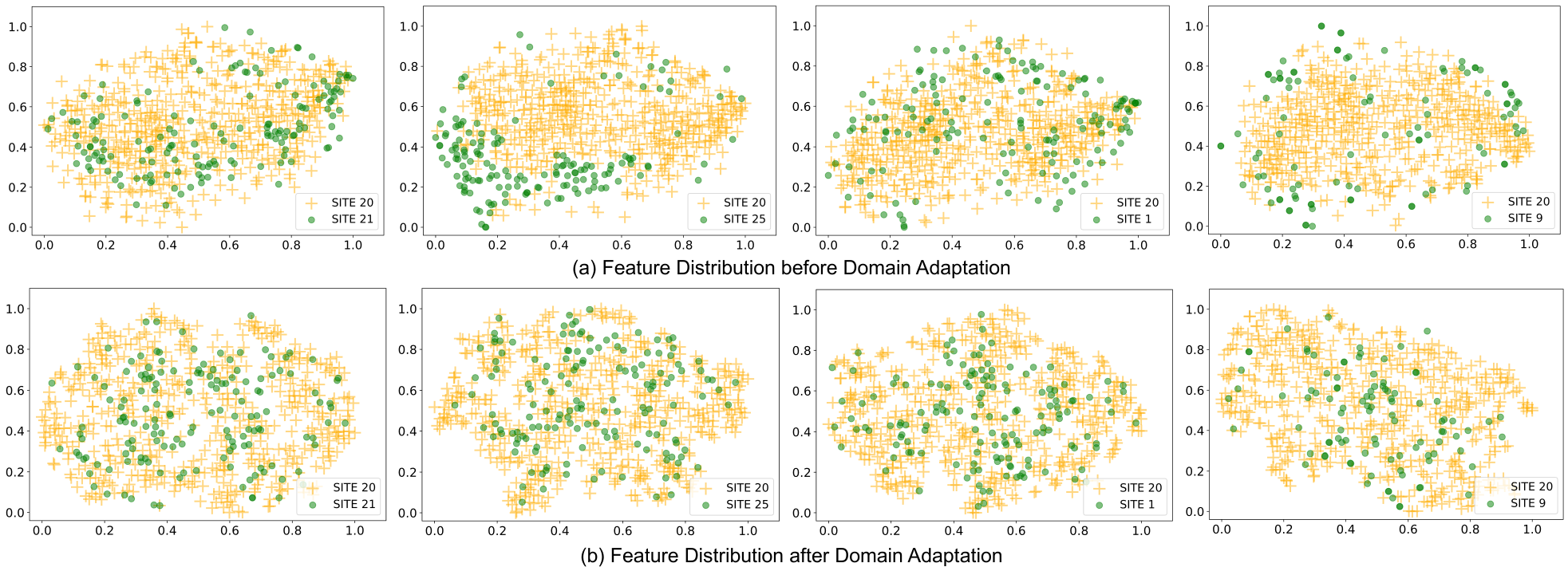}
	\caption{Visualization of feature distributions at different sites before and after domain adaptation by t-SNE~\cite{van2008visualizing}, where yellow "+" denotes the source domain and green dots denote the target domains.}
	\label{VIS}
\end{figure*}  

\subsection{Competition Methods}
We compare the proposed AUFA with three traditional (BC-SVM, LE-SVM and MNSFS) and four deep learning methods (GIN, ST-GCN, CORAL, and DANN).
Among them, BC-SVM, LE-SVM, MNSFS, GIN and ST-GCN have not undergone adaptive learning, meaning that the model trained on the source domain is directly used to predict category labels of target data. 
Note that CORAL, DANN and AUFA are different domain adaptation methods, and unlabeled target domain data 
will be involved in the training of the model.

(1) \textbf{BC-SVM}:
In this method, an FCN is constructed for each subject based on PC.
We then extract the betweenness centrality (BC)~\cite{freeman2002centrality} of each ROI (node) from the FCN to measure the number of shortest paths between that ROI and other ROIs in the brain.
Finally, the corresponding FCN for each subject has a 116-dimensional feature and is classified by the SVM.

(2) \textbf{LE-SVM}:
Similar to BC-SVM, we extract the local efficiency~\cite{achard2007efficiency} of each node as features ($116$ dimension) based on the constructed FCN, which is then utilized for classification through SVM.

(3) \textbf{MNSFS}:
The method extracts multiple measures for each node from the estimated FCN, including local efficiency, three different definitions of local clustering coefficients, and four centralities. 
The measures belonging to each node are assigned to a group, followed by sparse group LASSO for feature selection and SVM for classification~\cite{zhang2022selecting}.

(4) \textbf{GIN}:
It uses graph isomorphic network (GIN)~\cite{xu2018powerful} for FCN feature extraction. 
Each GIN layer aggregates the neighbor features of each node based on FCN and uses MLP to perform nonlinear transformation to achieve node update. 
This process can be repeated through multiple layers of the same structure to gradually fuse the global information.
Finally, the output of the GIN module is average pooled to obtain the graph representation features for each subject, and classify them by the fully connected layer. 
The method includes $2$ layers of GINs, with each layer having an output dimension of $64$.

(5) \textbf{ST-GCN}:
It models dynamic features of fMRI through spatio-temporal graph convolutional networks ~\cite{gadgil2020spatio}. 
We input regional fMRI time series into two spatio-temporal graph convolutional (ST-GC) units to learn the spatio-temporal representation of fMRI. 
Finally, the classification tasks of MDD and NC are performed through fully connected layers. 
Each ST-GC unit outputs features with a dimension of $64$.

(6) \textbf{CORAL}:
In this method, we use the same graph representation learning method as the proposed AUFA to extract features from each subject's FCN. 
Here, we use the labels of the source domains to perform supervised training and apply the covariance alignment (CORAL)~\cite{sun2016deep} module to align the distributional differences between the domains, thus reducing the effect of domain offset.

(7) \textbf{DANN}:
The domain adversarial neural network (DANN)~\cite{ganin2016domain} is a deep learning model for domain adaptation. 
In DANN, a convolutional neural network is used to learn representations of the source/target domain, and the domain discriminator is used to judge which domain the representation comes from. 
Similar to AUFA, DANN only has supervision of source domain and learns domain invariant features through adversarial training of domain discriminators and classifiers.

\subsection{Experimental Results}
{\color{black}
The experimental results of our proposed AUFA and $7$ comparison methods in MDD identification tasks are reported in Table~\ref{results}.
Statistical significance analysis results are reported in \emph{Supplementary Materials}.
We can make the following observations from the Table~\ref{results}. 
\emph{First}, the methods employing domain adaptive strategies (\emph{i.e.}, CORAL, DANN, and our proposed AUFA) exhibit significant advantages when compared to learning methods without domain adaptive processing (\emph{i.e.}, BC-SVM, LE-SVM, MNSFS, GIN, and ST-GCN). 
This phenomenon reveals the limitation on the generalization performance of models trained directly from the source domain and tested directly in the target domain. It is attributed to the significant data heterogeneity issue across different sites.
This validates the effectiveness of utilizing the maximum mean difference constraint to achieve inter-domain feature alignment.
\emph{Second}, deep learning models such as GIN and ST-GCN show relatively high performance in methods that do not perform domain adaptation. This is likely due to their capability to adaptively learn discriminative features that are advantageous for label-supervised classification tasks.
In contrast, traditional machine learning methods heavily rely on predefined features and prior knowledge. 
\emph{Furthermore}, our AUFA model achieves excellent performance on most of the metrics in several domain adaptive methods.
Compared with CORAL, which adopts the same graph representation learning in our method, AUFA achieves an average improvement of about 5\% and 4\% in ACC and AUC metrics on the four domain adaptation tasks. This improvement can be attributed to the augmentation-based self-optimization module integrated within AUFA, which helps alleviate the problem of model overfitting on the source domain. 
\emph{Finally}, compared to the popular DANN method, AUFA still maintains outstanding performance with an average improvement of about 5.5\% and 5\% in ACC and AUC metrics across the four domain adaptation tasks. 
Besides the augmentation of target data by AUFA, which mitigates the risk of overfitting, another possible reason is that the convolutional neural network structure adopted by DANN may not fully recognize and focus on the importance of features from different brain regions.  
In contrast, AUFA leverages self-attention mechanisms to learn different weights for each brain region, thus extracting more discriminative features.
}

\subsection{Visualization of FMRI Data Distribution}
{\color{black}
We visualize the distribution differences between the features at each site before and after domain adaptation. 
Specifically, we flattened the upper triangular matrix of the FCNs of the source and target domain data as the features without domain adaptation, and then, the features learned by the Transformer in our AUFA framework as the features after domain adaptation. 
Finally, these features are embedded into a 2D space for visualization by t-SNE~\cite{van2008visualizing}, with results displayed in Fig.~\ref{VIS}.
As shown in Fig.~\ref{VIS}, the feature distributions of the two sites that have not undergone domain adaptation show a large domain offset. 
After domain adaptation, the distributions of corresponding sites in the source and target domains become consistent. 
This suggests that our AUFA can effectively mitigate the heterogeneity in fMRI data from different sites, thereby enhancing the model's performance on the target domain.
}

\section{Discussion}
\label{S6}
\subsection{Ablation Experiment}
{\color{black}
To explore the effectiveness of the AUFA framework more deeply, we design three experiments to further analyze the effect of each sub-module:
(1) Feature extraction is performed using the same backbone network as AUFA, but without any domain adaptation and augmentation operations (only $L_{C}$ is retained), \emph{i.e}., AUFA-C; 
(2) The domain adaptation module in the AUFA framework is discarded and only $L_{C}$ and $L_{A}$ are retained, \emph{i.e}., AUFA-AUG; 
(3) Without augmenting the data in the target domain, only $L_{C}$ and $L_{M}$ are retained, \emph{i.e}., AUFA-MMD. 
We conduct ablation experiments on the largest sample classification task (\emph{i.e}., SITE 20 $\rightarrow$ SITE 21), and the results are reported in Table~\ref{variants}. 
It can be observed that our AUFA obtains the best classification results, and removing any part of the framework results in a degradation of performance. 
This demonstrates the effectiveness of the AUFA framework, wherein the maximum mean difference constraint effectively achieves inter-domain feature alignment. Additionally, the augmentation of the target domain alleviates overfitting and improves model generalizability. 
}

\subsection{Impact of Hyperparameters}
{\color{black}
The proposed AUFA is jointly optimized by cross-entropy loss, maximum mean difference loss and augmentation-based self-optimization loss. 
To explore the influence of different components of the loss function on the model, we conduct experiments on the domain adaptation classification task on SITE 20 $\rightarrow$ SITE 21. 
We set the hyperparameters $\lambda_{1}$ and$\lambda_{2}$ within the range $\{0.01, 0.05, 0.1, 0.5, 1, 5, 10, 15, 20\}$. The results are shown in  Fig.~\ref{parameters}.
As shown in Fig.~\ref{parameters}, 
the accuracy of AUFA increases gradually as $\lambda_{1}$ rises, reaching its peak when $\lambda_{1}$ is $1$.
This demonstrates that the maximum mean difference loss can effectively align the inter-domain features, thus improving model performance. 
But when $\lambda_{1}$ takes too large a value, the performance of our method shows a significant decrease. The possible reason is that when $\lambda_{1}$ is excessively large, the optimization process ignores the supervisory loss from the source domain labels, leading to underfitting of the AUFA model.
In addition, the overall trend of model performance with $\lambda_{2}$ is consistent with $\lambda_{1}$, and AUFA performs best at $\lambda_{2}$ is 1. When $\lambda_{2}$ is too large, the optimization process will ignore the supervised loss of source domain labels, leading to degraded performance. 
When $\lambda_{2}$ is too small, the effect of reducing model overfitting will not be achieved. Therefore, the choice of appropriate hyperparameter values is crucial for model performance.
}

\begin{table}[!tbp]
\setlength{\belowcaptionskip}{-8pt}
\setlength{\abovecaptionskip}{-1pt}
\setlength\abovedisplayskip{-1pt}
\setlength\belowdisplayskip{-2pt}
\footnotesize
\renewcommand{\arraystretch}{1.1}
\centering
\caption{Experimental results of AUFA and its three variants in the MDD and NC classification task for cross-domain SITE 20 $\rightarrow$ SITE 21, with the best results highlighted in bold.}
	\scriptsize
	\centering
	\setlength{\tabcolsep}{0.6pt}
	\begin{center}
    \begin{tabular*}{0.48\textwidth}{@{\extracolsep{\fill}} l|c  ccc c}
			\toprule
		    ~ Method     & Accuracy  & Precision           & Recall           & AUC           & F1-score                                   \\
			\hline
           ~ AUFA-C      & 0.54$\pm$0.03 & 0.59$\pm$0.03 & 0.58$\pm$0.04  & 0.58$\pm$0.01  & 0.58$\pm$0.02 \\
           ~ AUFA-AUG       & 0.55$\pm$0.01      & 0.59$\pm$0.01  & 0.62$\pm$0.05 & 0.56$\pm$0.01  & 0.60$\pm$0.02  \\
           ~ AUFA-MMD      & 0.56$\pm$0.01      & 0.60$\pm$0.01 & 0.61$\pm$0.03 & 0.56$\pm$0.01 & 0.60$\pm$0.02 \\
            \hline
        ~ AUFA~(Ours)  & \textbf{0.63$\pm$0.03} & \textbf{0.66$\pm$0.03}  & \textbf{0.73$\pm$0.06}  & \textbf{0.62$\pm$0.04} & \textbf{0.69$\pm$0.03}   \\
			\bottomrule
		\end{tabular*}
	\end{center}
	\label{variants}
\end{table}

\subsection{Discriminative Brain Regions}
\begin{figure}[t]
	\setlength{\abovecaptionskip}{-1pt}
	\setlength{\belowcaptionskip}{-2pt}
	\setlength\abovedisplayskip{-1pt}
	\setlength\belowdisplayskip{-1pt}
	\centering
	\includegraphics[width=0.49\textwidth]{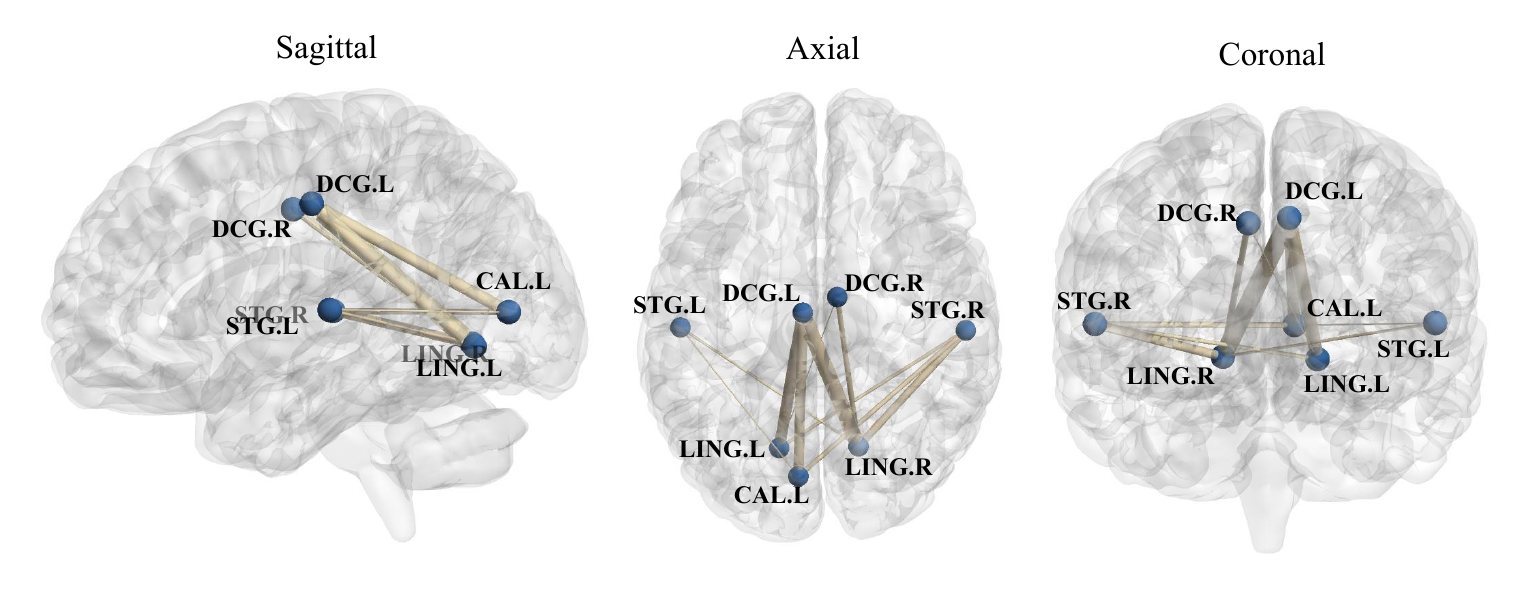}
	\caption{Visualization of the top ten most discriminative functional connections detected by AUFA in the MDD vs. NC classification task using the BrainNet Viewer~\cite{xia2013brainnet}.}
	\label{ROI}
\end{figure}

{\color{black}
This work addresses the problem of data heterogeneity in multi-site fMRI while identifying the most discriminative brain regions in the MDD vs. NC classification task. 
In Section~\ref{1}, 
we utilize self-attentive mechanisms to learn that the attention matrix $A$ can effectively quantify the interdependence between different brain regions.
Given that our proposed AUFA framework is used for brain disorder detection tasks, the elements with larger values in $A$ indicate those brain region connections that play a decisive role in distinguishing healthy from diseased states. 
To gain deeper insight into these critical connections, we averaged the multiple attentional matrices learned for each attention head, aiming to strengthen those significant connections consistently apparent across multiple attentional perspectives. 
Subsequently, the top ten most discriminative brain region connections were visualized by the BrainNet Viewer~\cite{xia2013brainnet} tool, and the results are presented in Fig.~\ref{ROI}.
As can be seen in Fig.~\ref{ROI}, the identified brain regions such as the calcarine fissure (CAL), superior temporal gyrus (STG), lingual gyrus (LING), and medial and cingulum gyrus (DCG) play important roles. 
This is highly correlated with previous studies of MDD~\cite{kang2017reduced,su2018regional,yang2016diminished,long2020altered},
thus validating the effectiveness of AUFA based on rs-fMRI in identifying disease-related functional connectivity patterns. 
In the \emph{Supplementary Materials}, we also elaborate on the influence of different source domains.
}

\begin{figure}[!t]
	\setlength{\abovecaptionskip}{-1pt}
	\setlength{\belowcaptionskip}{-2pt}
	\setlength\abovedisplayskip{-1pt}
	\setlength\belowdisplayskip{-1pt}
	\centering
	\includegraphics[width=0.48\textwidth]{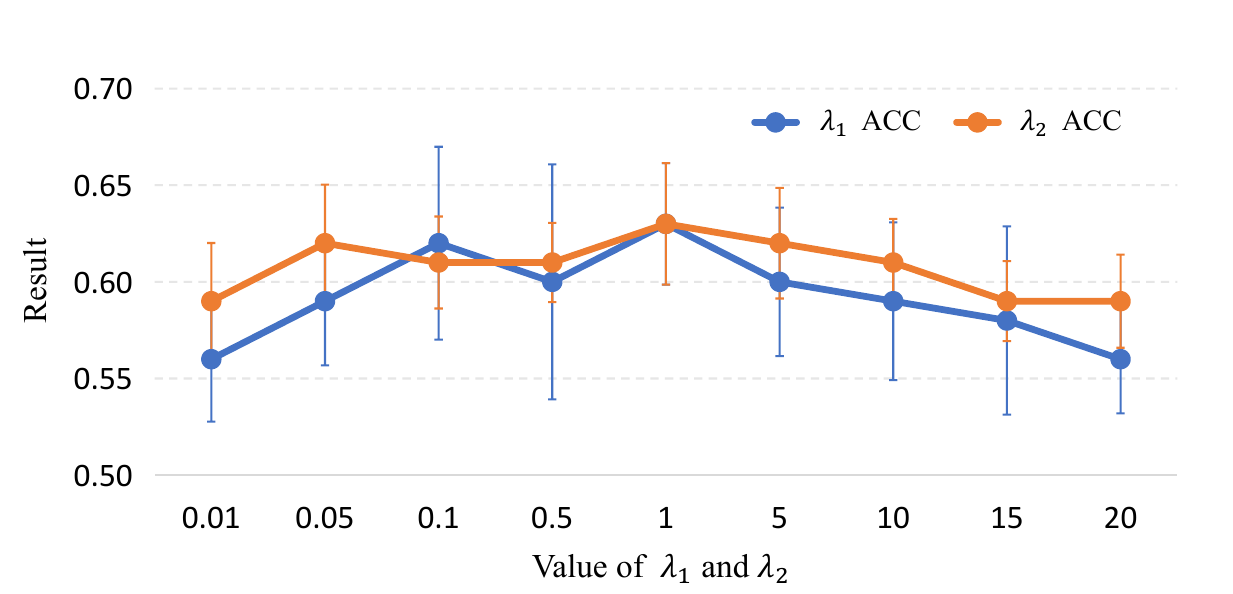}
	\caption{Accuracy (ACC) achieved by AUFA in MDD and NC classification with different values of $\lambda_{1}$ and $\lambda_{2}$.}
	\label{parameters}
\end{figure}

\subsection{Limitations and Future Work}
{\color{black}
There are still several limitations to the present study.
\emph{First}, 
The proposed framework only considers resting-state fMRI data, while there exists untapped potential in integrating other modalities like structural MRI and demographic information. These additional sources offer valuable, diverse insights into the brain, enriching our understanding from multiple viewpoints.
We will employ multi-modal data in the future to help boost learning performance.
\emph{Second},
this work mainly focuses on dealing with the migration problem of a single source/target domain.
We can develop more advanced models to integrate and optimize information from multiple source domains, and explore effective adaptive strategies for complex scenarios involving multiple source/target domains.
\emph{In addition}, it is interesting to generalize the model to unknown domains. 
In the future, we will focus on constructing and optimizing generative models for modeling data differences between source and target domains.
This will facilitate the migration of knowledge, thereby enhancing the overall generalization performance of our approach across unknown domains.
}

\section{Conclusion}
\label{S7}
{\color{black}
In this paper, we construct an augmentation-based unsupervised cross-domain fMRI adaptive framework for the automatic diagnosis of major depressive
disorder (MDD). 
We aim to reduce the distributional differences of fMRI data from different sites and address the overfitting of the model in the source domain.
We learn graph-level features with spatial attention via Transformer encoder and mitigate data heterogeneity between domains via the domain adaptation module. 
We also augment the target domain data to 
enhance the model's generalization in that specific domain.
Experimental results on multiple sites validate the effectiveness of AUFA in MDD detection. 
In addition, our framework can discover discriminative brain regions,  providing potential imaging biomarkers for MDD analysis.
}

{\color{black}
\footnotesize
\bibliographystyle{IEEEtran}
\bibliography{References}
}

\end{document}